\documentclass[aps,prl,twocolumn,groupedaddress,10pt]{revtex4-1}

\usepackage{graphicx}
\usepackage{epsfig} %

\begin{document}

\title{Sub-diffraction-limited quantum imaging within a living cell}


\author{Michael~A.~Taylor,$^{1,2}$ Jiri~Janousek,$^{3}$ Vincent~Daria$^{4}$, Joachim~Knittel$^{1}$,\\
 Boris~Hage$^{3,5}$, Hans-A.~Bachor$^{3,4}$, and Warwick~P.~Bowen$^{2\ast}$}

\affiliation{$^1$Department of Physics, University of Queensland, St Lucia, Queensland 4072, Australia}
\affiliation{$^2$Centre for Engineered Quantum Systems, University of Queensland, St Lucia, Queensland 4072, Australia}
\affiliation{$^3$Department of Quantum Science, Australian National University, Canberra, ACT 0200, Australia}
\affiliation{$^4$John Curtin School of Medical Research, Australian National University, Canberra, ACT 0200, Australia}
\affiliation{$^5$Institut f\"{u}r Physik, Universit\"{a}t Rostock, D-18051 Rostock, Germany}
\affiliation{$^\ast$e-mail:  wbowen@physics.uq.edu.au}

\begin{abstract}
 We report both sub-diffraction-limited quantum metrology and quantum enhanced spatial resolution for the first time in a biological context. Nanoparticles are tracked with quantum correlated light as they diffuse through an extended region of a living cell in a quantum enhanced photonic force microscope. This allows spatial structure within the cell to be mapped at length scales down to 10~nm.  Control experiments in water show a 14\% resolution enhancement compared to experiments with coherent light. Our results confirm the longstanding prediction that quantum correlated light can enhance spatial resolution at the nanoscale and in biology. Combined with state-of-the-art quantum light sources, this technique provides a path towards an order of magnitude improvement in resolution over similar classical imaging techniques.
\end{abstract}

\maketitle


The emerging field of quantum imaging utilizes quantum effects to overcome classical imaging constraints. In particular, non-classical states of light can allow the shot-noise and diffraction limits to be surpassed~\cite{Kolobov2000}, and quantum engineered artificial atoms allow new approaches to sensing~\cite{McGuinness2011}. The primary motivation for such techniques is in biological imaging~\cite{McGuinness2011,Treps2002,Nasr2009,Ono2013,Brida2010}, where any improvement in imaging technology can reveal new levels of cellular complexity. Since sub-cellular structures often have nanometer size scales, spatial resolution surpassing the diffraction limit is particularly beneficial. However, neither sub-diffraction-limited resolution nor quantum enhanced resolution have previously been achieved in biology. To date, the only reported demonstration of biological imaging with non-classical light has been in  dispersion compensation for optical coherence tomography~\cite{Nasr2009}. Even in non-biological demonstrations, both the absolute sensitivity and resolution of optical quantum imaging has been constrained to levels far inferior to state-of-the-art classical technology~\cite{Ono2013,Brida2010,Pittman1995}. While unprecedented sensitivity is in principle achievable using squeezed states of light~\cite{Kolobov2000,Treps2002}, no method has been experimentally demonstrated capable of utilizing squeezed light in biological imaging.

 Here we propose and demonstrate a new quantum imaging method which applies squeezed light in photonic force microscopy (PFM)~\cite{Florin1997}. This allows both quantum enhanced resolution and sub-diffraction-limited quantum imaging in biology for the first time, with resolution comparable to leading classical experiments.  PFM is a classical sub-diffraction-limited imaging technique closely analogous to atomic force microscopy (AFM), but with a nanoscale particle trapped in optical tweezers replacing the probe tip~\cite{Florin1997,Ghislain1993}. As the nanoparticle explores a cell, environmental variations which affect its thermal diffusion can be mapped.  PFM has been used to map both 3D surfaces~\cite{Friese1999,Tischer2001} and mechanical properties of fluids~\cite{Pralle1998}, and has been applied to study cell membranes~\cite{Florin1997}, nanoscale protein motors~\cite{Scholz2005}, molecular interactions~\cite{Rohrbach2004}, and, similar to our work here, intracellular viscoelasticity~\cite{Bertseva2009}. As is typical of nanoprobe based microscopy techniques, the spatial resolution achievable in PFM  is not constrained by the diffraction limit. 
The resolution lateral to the motion of the probe particle is constrained by its size. However, similar to AFM, the resolution along the direction of motion is typically limited by measurement signal-to-noise~\cite{Friese1999,Rohrbach2004}.
 Here we use non-classical light to improve the signal-to-noise, and thereby demonstrate quantum enhanced resolution in PFM.

The squeezed-light-enhanced PFM is used to construct one-dimensional profiles of spatial structures within a cell, with features observed at length scales down to 10 nm. Control measurements in water confirm that for fixed optical power, squeezed light provides 14\% enhancement over the resolution possible with coherent light. A 74\% increase in optical power would be required to achieve this level of enhancement without squeezing; increasing the potential for damage~\cite{Peterman2003,Neuman1999} and photochemical disruption of cellular processes~\cite{Neuman1999,Lubart2006}, which are known to severely limit biological applications of PFM~\cite{Rohrbach2004}. By demonstrating for the first time that non-classical light can improve resolution in a biological context, the PFM achieves the key requirement for quantum enhanced imaging in biology. Our results, further, constitute the first demonstration of quantum enhanced resolution using squeezed light in any context. When combined with 3D particle tracking, quantum enhanced nanoscale images of biological structure could be constructed, placing practical applications of quantum imaging with non-classical light within reach.

The results presented here complement previous quantum imaging experiments using non-classical light. In imaging applications that simultaneously sample the entire field of view, many spatial modes are captured. Quantum enhancement then requires that quantum correlations are established between a large number of these modes. Such multimode non-classical light has been applied in proof-of-principle demonstrations of sub-shot noise absorption imaging~\cite{Brida2010}, enhanced 2-photon microscopy~\cite{Fei1997}, ghost imaging via photon correlations~\cite{Pittman1995}, improved image reconstruction against a noisy background~\cite{Lopaeva2013}, generation of entangled images~\cite{Boyer2008}, noiseless image amplification~\cite{Mosset2005,Lopez2008}, and to eliminate unwanted artifacts in optical coherence tomography~\cite{Nasr2009}.  However, practical applications have been precluded by a lack of both bright multi-mode sources of strongly non-classical light, and high bandwidth array detectors capable of efficiently detecting this light~\cite{Lassen2007}. By contrast, in single or few-modes scenarios, such sources and detectors are readily available~\cite{Armstrong2012}. This has enabled quantum enhanced measurements of spatial parameters such as laser beam deflection~\cite{Treps2002,Treps2003} and spatial quantum correlations~\cite{Janousek2009}. However, these experiments suffer the apparent major drawback that quantum enhancement is only possible for a number of pixels at most equal to the number of available single-mode sources of quantum correlated light. Consequently, they have previously been limited to a maximum of 8 pixels~\cite{Armstrong2012}. This limitation can be overcome using a scanning probe as demonstrated here for the first time, or an optical raster scan as recently demonstrated in Ref.~\cite{Ono2013}.



  \begin{figure}
 \begin{center}
   \includegraphics[width=8.6cm]{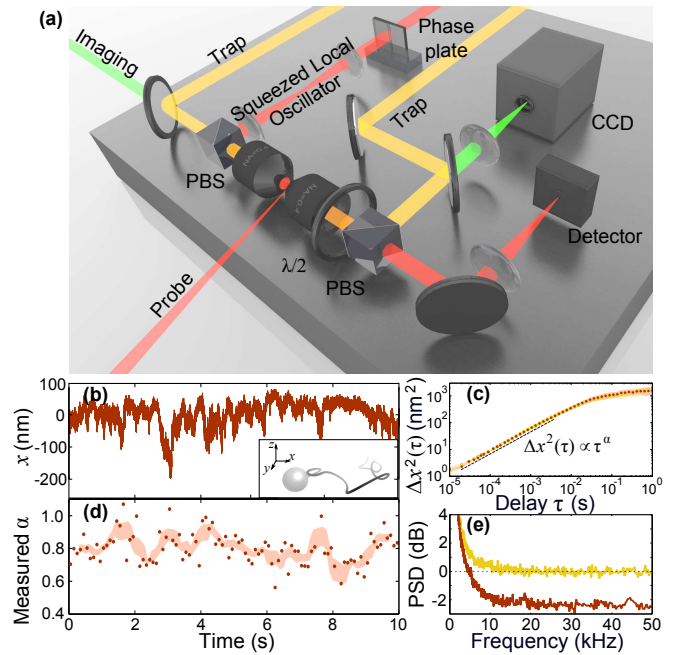}
   \caption{  Experimental setup. 
 (a), Counter-propagating trapping fields (orange) confine particles between two objectives, and are isolated from the detector with polarizing beamsplitters (PBS) and waveplates ($\lambda/2$). An imaging field (green)  allows visual identification of the particles near the optical trap on a CCD camera. The particle tracking measurement relies only on an amplitude squeezed local oscillator and an amplitude modulated probe (red), with the probe providing dark-field illumination, and the particle tracking signal arising from interference between scattered light from the probe and the local oscillator. (b), Measured particle motion, which is the $x$ projection of the 3D motion (shown schematically in the inset). (c), The MSD is constructed with both squeezed light (dark red) and coherent light (gold), and $\alpha$ determined by fitting this to Eq.~\ref{alpha_Def}. The classical and squeezed example traces here both yield $\alpha=0.83$. (d), The raw data was divided into 100~ms segments and the value of $\alpha$ established for each (solid dots). The light red shaded region represents the moving mean and standard error with a 0.5 second width.  (e), The normalized power spectral density (PSD) shows that squeezing suppressed the noise floor by 2.4~dB.
}
 \label{LayoutData}  
 \end{center}
\end{figure}

In PFM, a probe particle is tracked as it is scanned over the field of view. Variations in measured motion are then studied over a measurement time which is often of order minutes~\cite{Friese1999,Tischer2001}. Provided the microscope has sufficient stability~\cite{Rohrbach2004}, and that these variations are dominated by static intracellular structure, rather than dynamic cellular processes~\cite{Selhuber2009} or nanoscale motion of cellular constituents, this allows a map of the cellular structure to be constructed. The quantum PFM reported in this Article utilizes a recently developed quantum enhanced nanoparticle tracking technique~\cite{Taylor2013_sqz}. While Ref.~\cite{Taylor2013_sqz} provided a tool to study temporal fluctuations within living cells, the lack of spatial resolution was a critical shortcoming, preventing any conclusions from being drawn regarding the dominant source of fluctuations in the measured motion, and therefore application as a PFM. Here, spatial resolution is introduced and quantum PFM realized, with thermally driven motion used to scan the probe particle through an extended region of the cell~\cite{Tischer2001}.

 The experimental setup (shown in Fig.~\ref{LayoutData}(a)) features several important modifications from a conventional PFM, facilitating the use of amplitude squeezed light to enhance measurement sensitivity. Because most biological processes occur at Hz-–kHz frequencies, where classical noise sources constrain the possibility of generating squeezing~\cite{McKenzie2004}, a novel optical lock-in technique~\cite{Taylor2013_lockin} is used to evade low frequency noise and allow quantum enhancement at these frequencies. This is combined with dark-field illumination to remove unwanted light from the measured optical field~\cite{Kudo1990,Taylor2013darkfield}. Additionally, self-homodyne measurement is used, which allows the local oscillator itself to be squeezed, such that the squeezed field perfectly overlaps with the local oscillator at detection even without prior knowledge of the spatial mode shape after propagation through the living cell and high numerical aperture lenses.

 {\em Saccharomyces cerevisiae} yeast cells were immobilized with an optical trap, and lipid granules of approximately 300~nm diameter were tracked with either squeezed or coherent light as they diffused within the cellular cytoplasm. Since the characteristic thermal motion of any particle is determined by the mechanical properties of its surrounding medium, such intracellular particle tracking measurements are commonly used to study the mechanics of cellular cytoplasm~\cite{Bertseva2009}. Lipid granules are well suited for use as probe particles within yeast as they occur naturally and can be tracked precisely due to their high refractive index~\cite{Norrelykke2004,Selhuber2009}. In our experiment, the high frequency thermal motion of a lipid granule reveals the viscoelastic mechanical properties of the surrounding cytoplasm, while for sufficiently long measurements its slow thermal drift provides spatial resolution by bringing it into contact with  different parts of the cellular cytoplasm. Thus, spatial inhomogeneity in the viscoelasticity can be quantified by a single continuous measurement of the lipid granule position. In our experiment, the particle position $x(t)$ transverse from the trap center was measured by combining scattered light from the sample with a  local oscillator field which was spatially shaped such that direct measurement of the total power yielded the  particle position (Fig.~\ref{LayoutData}(b)).  In the same manner as in Ref.~\cite{Taylor2013_sqz}, mechanical properties of the cytoplasm directly surrounding the nanoparticle could be characterized from its mean squared displacement (MSD) after a delay $\tau$, 
 \begin{equation} 
\left< \Delta x^2(\tau) \right>  = \left< \left( x(t)-x(t-\tau) \right)^2 \right>, \label{MSD_def}
\end{equation}
with an example shown in Fig.~\ref{LayoutData}(c).  Squeezed light improves the precision by reducing the error with which the MSD can be estimated. This improvement is shown in the measured power spectral density  (Fig.~\ref{LayoutData}(e)), with squeezed light lowering the noise floor by 2.4~dB.  For short delays, the MSD is dominated by thermal motion and has the form
  \begin{equation}
\left< \Delta x^2(\tau) \right> =  2 D \tau^\alpha ,\label{alpha_Def}
\end{equation}
 where the diffusive parameter $\alpha$ carries information about the mechanical properties of the surrounding medium~\cite{Gittes1997,Mason1997}. $\alpha$ is determined for a set of data by fitting the MSD at short delays to Eq.~\ref{alpha_Def}.  When $\alpha=1$, the motion is diffusive, which is indicative of a random walk type of motion, whereas confinement of the particle causes subdiffusive motion ($0<\alpha<1$). Subdiffusive motion is an indicator that the cellular cytoplasm exhibits both viscosity and elasticity~\cite{Gittes1997}, since to constrict motion the cytoplasm must store mechanical energy. In our experiments, 100~ms of data was sufficient to precisely determine $\alpha$. Consequently, the measured values of $\alpha$ allowed temporal variations in the cellular viscoelasticity to be characterized with 10~Hz bandwidth.


As lipid particles undergo three dimensional (3D) thermal motion, they are exposed to different parts of the cell (Fig.~\ref{LayoutData}(b) inset).  In a full PFM,  3D motion is tracked through an extended region of the cell. By characterizing the changes in $\alpha$ that occur, it is then possible to construct 3D images of structure within the cellular cytoplasm~\cite{Bertseva2009}. Here, to demonstrate that non-classical light enables resolution surpassing that possible with coherent light, a proof-of-principle demonstration is achieved using 1D particle tracking along the $x$ axis,  with the co-ordinates $y$ and $z$ not determined. This allows 1D profiles of $\alpha(x)$ to be constructed following the projection of the trajectory onto the $x$ axis.

 \begin{figure*}
 \begin{center}
   \includegraphics[width=17.5cm]{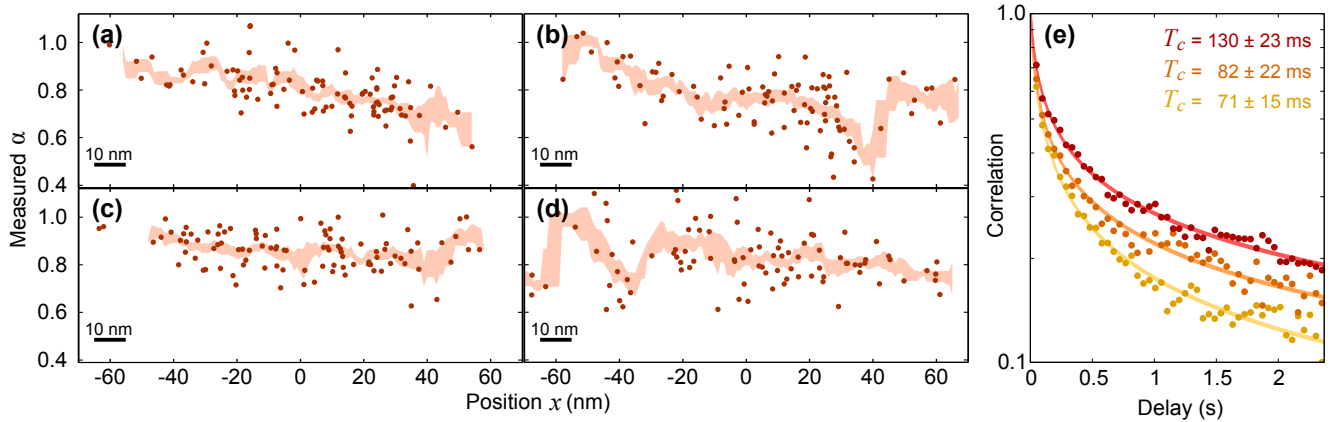}
   \caption{  1D profiles of $\alpha$. 
 Each circle represents a single measurement of $\alpha$ vs $x$ using a 100~ms set of data. The shaded regions represent the running mean and standard error with 10~nm resolution (thick black bar). Each profile was recorded minutes apart to allow the particle time to diffuse to different regions of the cell with qualitatively different spatial structures.  The particle confinement is greatest where $\alpha$ is lowest, such as the dip about 40~nm in (b), and the particle movement is most free when $\alpha$ is highest, such as the peak at -55~nm in (d).  To verify that the changes in $\alpha$ are spatial, correlations between sequential measurements of $\alpha(x)$ are analyzed  for three sets of data in (e). The circles are experimentally determined correlations between a series of measurements, which are well fitted by the predicted relation (lines) explained in Supplementary section S1. By fitting data to this theory, the characteristic time $T_c$ for the particle to diffuse into uncorrelated regions of the cell can be determined. 
Note that the decay in correlation restricts the duration over which $\alpha(x)$ profiles can be constructed.
Although the characteristic times found here are in the range of 0.1~s, correlations were found to persist for sufficient time to construct the 10~s $\alpha(x)$ profiles shown here.
   }
 \label{AlphaX} 
 \end{center}
\end{figure*}

 A series of experiments were performed in which the motion of lipid particles was tracked with quantum enhanced precision for 10~s as they diffused through the cell. The data from each experiment was separated into 100~ms segments, with both $\alpha$ and the mean position along the $x$ axis determined for each segment. As the particle diffused, a profile of $\alpha$ was generated as a function of $x$, with four representative profiles shown in Fig.~\ref{AlphaX}.  
 As can be seen, the particles explored a range of approximately 120~nm along the $x$ axis over the 10~s measurement interval, consistent with the MSD in Fig.~\ref{LayoutData}(c) extrapolated to longer delays. 
The directly obtained data exhibited substantial noise both from the measurement process and due to the unknown trajectory of the particle in the $y$ and $z$ directions.  To identify statistically resolvable features, the running mean and standard error of $\alpha$ were calculated along the $x$ axis, with a 10~nm averaging window defining the spatial resolution. Since increased spatial averaging makes small changes in $\alpha$ easier to resolve, an intrinsic compromise is present between  spatial resolution and contrast, with the latter defined as the statistical uncertainty in $\alpha$. The choice of 10~nm spatial resolution was found to provide sufficient contrast to observe cellular structure.

 The observed spatial structure varies between measurements of $\alpha(x)$ because the particle follows different 3D trajectories. Gradual linear changes in $\alpha$ were observed (e.g. Fig.~\ref{AlphaX}(a)) which suggest a spatial gradient in the molecular crowding along the $x$ axis~\cite{Weiss2004crowding}, along with narrow dips in $\alpha$ (e.g. Fig.~\ref{AlphaX}(b) at 40~nm) suggestive of barriers in the cytoplasm, areas of homogeneity (e.g. Fig.~\ref{AlphaX}(c)), and peaks in $\alpha$ (e.g. Fig.~\ref{AlphaX}(d) at -55~nm) which may follow from small voids in the cytoplasmic structure. Since only the projection of the particle motion onto the $x$ axis was tracked, it is not possible to define the complete trajectory along which these 1D profiles are taken. This obscures the biological origin of observed features. For instance, the narrow dip in $\alpha$ seen in Fig.~\ref{AlphaX}(b) could result from a range of subcellular components including an actin filament or the edge of a larger organelle, even though these have markedly different 3D profiles. 
The 3D motion of the particle also degrades the contrast of narrow features by averaging measurements of $\alpha$ from a range of positions along the $y$ and $z$ axes. 
 These limitations could be resolved by incorporating our technique in a 3D PFM~\cite{Florin1997} which maps the complete trajectory of the particle. It may then be possible to generate a quantum enhanced 3D image of the cell, with quantum enhancement only required for one axis  from which $\alpha$ could be determined.  

In addition to this technological improvement, it will also be important to minimize non-ideal effects in future quantum PFM systems. It is particularly important that the effect of  background scattering centers on the measured signal be characterized, and unwanted scattered light eliminated as much as possible (see Supplementary section S3). This technical difficulty has already been addressed in previous classical  measurements of anomalous diffusion within cells~\cite{Norrelykke2004}. While it requires careful attention, it does not present a fundamental barrier to use of this quantum imaging technique in practical biological experiments.

Importantly, even though the biological origin of the profiles in Fig~\ref{AlphaX}(a)--(d) is obscured, the measured changes in $\alpha$ can be rigorously shown to originate from spatial structure within the cell. 
 In a static spatially varying environment, spatial correlations between $\alpha(x)$ profiles should decay exponentially with the time between the profiles, as the unknown motion along the $y$ and $z$ axes brings the particle into different regions of the cell.
 By contrast, temporal changes in the cell~\cite{Selhuber2009} or drifts in the apparatus which produce fluctuations in $\alpha$ should not exhibit any correlations between profiles. A sequence of $\alpha(x)$ profiles were measured at a rate of 20~s$^{-1}$, and correlations between the profiles calculated. The experimental data (Fig.~\ref{AlphaX}(e)) shows excellent agreement with the predicted correlation decay for measurement in a static spatially varying environment (Supplementary section S1), thus confirming that the $\alpha$ profiles reflect spatial structure rather than temporal fluctuations. Furthermore, this analysis of correlations allows the local length scale of viscoelastic structure to be determined in the region of the nanoparticle. This length scale can be found by combining the MSD measured in Fig.~\ref{LayoutData}(c) with the fitted characteristic time $T_c$ over which the particle takes diffuses into an uncorrelated region of the cell (see Supplementary section S1). The length scales of the viscoelastic structure in three different regions were determined to be $46.9\pm1.3$~nm, $43.7\pm2.4$~nm, and $42.6\pm1.9$~nm (Fig.~\ref{AlphaX}(e)), demonstrating that changes in the characteristic length of spatial structure in different parts of the cell can be statistically distinguished with nanometer precision.



\begin{figure*}
 \begin{center}
   \includegraphics[width=15cm]{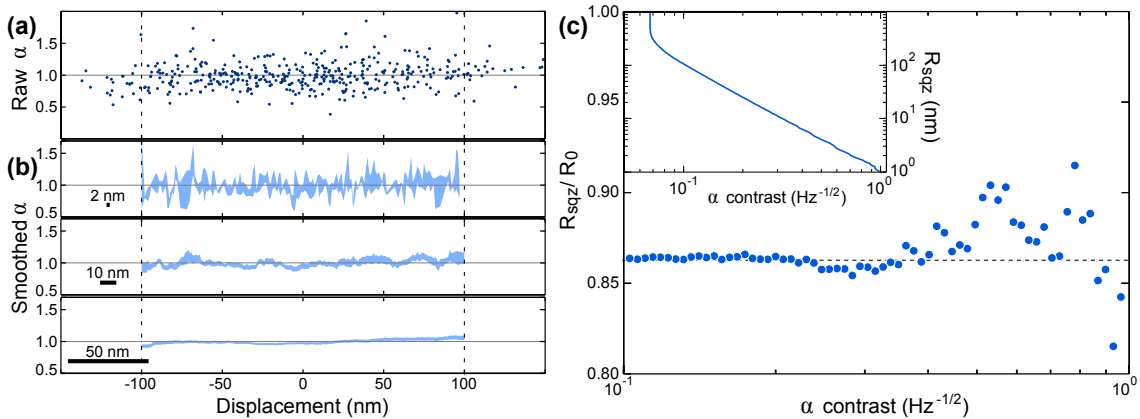}
   \caption{ Characterization of the spatial resolution.
 To calibrate the resolution enhancement achieved here, a profile of $\alpha(x)$ was constructed by tracking particles in water. The individual measurements are shown in (a), while (b) shows the corresponding moving mean and standard error calculated over a 200~nm range with spatial resolution of 2~nm, 10~nm, and 50~nm. The data closely follows the expected  $\alpha=1$ result (horizontal line).  (c), The ratio of spatial resolutions with squeezed and coherent light is shown as a function of contrast in $\alpha$. The $\alpha$ contrast values shown  are normalized into units of Hz$^{-1/2}$ to account for the improvement in absolute contrast as data is accumulated. For a fixed contrast, the spatial resolution achievable with squeezing (R$_{\rm sqz}$) is improved by approximately 14\% when compared to coherent light (R$_{0}$). The absolute spatial resolution achievable using squeezed light is plotted in the inset. Since the number of points being averaged is proportional to the spatial resolution, the sensitivity scales as the inverse square root of spatial resolution until the averaging window width becomes comparable to the spatial range of the measured data. 
   }
 \label{Resolution} 
 \end{center}
\end{figure*}

 Due to the complexity of the intracellular environment, the quantum resolution enhancement achieved in the PFM was characterized via control experiments on  1~$\mu$m radius silica beads in water, rather than directly {\it in vivo}. This approach of using simple well understood control experiments is standard when calibrating resolution in PFM~\cite{Florin1997,Friese1999,Rohrbach2004}. 
 $\alpha(x)$ profiles were constructed  from 80~s of data  using both squeezed and coherent light. In this case, 2.5~dB of squeezing was measured, which closely approaches the enhancement achieved in biological measurements. These measurements show no statistically significant spatial structure (Fig.~\ref{Resolution}), with $\alpha=1$ at all spatial locations as expected for Brownian motion.  This lack of statistically significant variation provides further verification that the structure observed {\it in vivo} can be attributed to changes in $\alpha$, rather than drifts in the apparatus. 
 Since water is homogeneous with $\alpha=1$ throughout, the variation in this data allows the statistical uncertainty, or contrast, of our measurements of $\alpha$ to be determined. 
The $\alpha$ contrast was characterized as a function of spatial resolution by varying the width of the running average along $x$, as shown representatively in Fig.~\ref{Resolution}(b).  As the width increases, the spatial resolution is degraded, but the contrast in $\alpha$  improves since more data is averaged at each position along $x$ (see Fig.~\ref{Resolution}(c) inset). 
By comparing the resolution required to achieve a fixed contrast with and without squeezed light, it was possible to confirm for the first time that squeezed light can be used to enhance spatial resolution. Furthermore, this also provides the first demonstration of quantum enhanced spatial resolution in a biological context. As shown in Fig.~\ref{Resolution}(c), in this proof-of-principle experiment squeezed light allowed a 14\% improvement in resolution for contrasts in $\alpha$ ranging from 0.1--1~Hz$^{-1/2}$.

 To take a specific example, the biological profiles shown in Fig.~\ref{AlphaX} use 10~nm resolution and include 10~s of accumulated data. In our control experiments in water, this resolution and accumulation time would allow structures that alter $\alpha$ by 0.1 to be resolved (see Fig.~\ref{Resolution}(c) inset).  By comparison, a resolution of 12~nm would be required to resolve such features with coherent light. 
 It is important to note that both the measurement sensitivity and the spatial range of diffusion differ between this calibration with silica beads and the measurements in biology. Consequently, the $\alpha$ contrast determined here differs from that achieved {\it in vivo}, where the average standard error in $\alpha$ was 0.04. 
 Importantly, the quantum resolution enhancement is independent of the absolute level of contrast, as can be seen over an order of magnitude in Fig.~\ref{Resolution}(c). Therefore, even though the absolute resolution differs between {\it in vivo} and control experiments, the quantum resolution enhancement predicted here can be expected to accurately represent the {\it in vivo} enhancement.

 In absolute terms, the resolution achieved here is comparable to that of leading classical PFM measurements of viscoelasticity~\cite{Scholz2005,Bertseva2009}. Furthermore, the resolution could be substantially improved using an increased level of squeezing.
With 10~dB of measured squeezing, as reported in a number of experiments~\cite{Stefszky2012,Mehmet2012}, an order of magnitude enhancement should be feasible (see Supplementary section S2). This could potentially allow Angstr\"{o}m level resolution.
 In principle, further enhancement may be possible by using more sophisticated quantum measurements~\cite{Taylor2013QNL,Tsang2009}, with recent theoretical results predicting that an array of photon number resolving detectors could even allow particle tracking at the de Broglie limit~\cite{Tsang2009}.

 When combined with the advances described above, the technology introduced here could help to answer important questions related to the nanoscale structure within cells. It has potential which extends beyond mapping of organelle positions, since thermal motion is critical to the operation of the cell and mediates important functions such as chemical reactions~\cite{Guigas2008} and protein folding~\cite{Frauenfelder2006}. It has been shown that the optimal diffusive regime is different for storage, transport, and chemical reactions~\cite{Guigas2008}, and that in some regions of the cell, structures which influence diffusion are organized at the nanoscale~\cite{Tseng2004}. It remains unknown to what extent these nanoscale structural variations reflect an underlying biological function. We anticipate that in the future, quantum imaging could play an important role in answering such questions.

 Our results complement recent biological applications of quantum engineered diamond probes with Nitrogen Vacancy (NV) centres, which have enabled thermal~\cite{Kucsko2013} and magnetic cellular imaging~\cite{LeSage2013,Steinert2013,Kaufmann2013}. None of these applications have achieved sub-diffraction limited resolution, as they rely on optically resolvable arrays of stationary NV probes~\cite{Kucsko2013,LeSage2013,Steinert2013,Kaufmann2013}, confining them to the study of relatively large cellular structures and organelles.  The resolution achieved here is over an order of magnitude finer, providing the possibility to observe important nanoscale cellular structures such as  membranes, actin networks, and individual proteins. Since the approach is in principle transferable to NV nanodiamond based imaging, it could also open the door to simultaneous sub-diffraction-limited imaging of structure, temperature and magnetic fields.

In summary, we report the first application of quantum imaging techniques to  sub-diffraction-limited biological imaging, and demonstrate that non-classical light can improve spatial resolution in biological applications.  The viscoelastic structure within a living yeast cell is sampled along the trajectory of a thermally driven nanoparticle, revealing spatial structure with length scales down to 10~nm.  Control experiments in water show that  the spatial resolution is enhanced by 14\% through use of squeezed light. Future experiments which apply this quantum enhanced photonic force microscope with improved technology may enable resolution of sub-nm structure {\it in vivo}.
\\

 This work was supported by the Australian Research Council Discovery Project Contract No. DP0985078 and the Australian Research Council Centre of Excellence for Engineered Quantum Systems CE110001013.





\begin{thebibliography}{10}


\bibitem{Kolobov2000}
M. I. Kolobov and C. Fabre, 
{\em Quantum limits on optical resolution,} 
  Phys. Rev. Lett. {\bf 85}, 3789--3792 (2000). 



















\bibitem{McGuinness2011}
L.~P. McGuinness {\em et al.},
{\em Quantum measurement and orientation tracking of fluorescent nanodiamonds inside living cells,}
  Nature Nanotech. {\bf 6}, 358--363 (2011).



\bibitem{Treps2002}
N. Treps, U. Andersen, B. Buchler, P.~K. Lam, A. Ma\^{i}tre, H.-A. Bachor, and C. Fabre, 
{\em Surpassing the standard quantum limit for optical imaging using nonclassical multimode light,} 
  Phys. Rev. Lett. { \bf 88}, 203601  (2002).

\bibitem{Nasr2009}
M. B. Nasr,  D. P. Goode, N. Nguyen, G. Rong, L. Yang, B. M. Reinhard, B. E. A. Saleh, and M. C. Teich, 
 {\em Quantum optical coherence tomography of a biological sample,}
 Opt. Commun. { \bf 282}, 1154--1159 (2009).






\bibitem{Ono2013}
T. Ono, R. Okamoto, and S. Takeuchi,  
{\em An entanglement-enhanced microscope,}
 Nat. Commun. {\bf 4}, 2426 (2013).


\bibitem{Brida2010}
G. Brida,  M. Genovese, and I. Ruo Berchera, 
{\em Experimental realization of sub-shot-noise quantum imaging,} 
  Nat. Photon. { \bf 4}, 227--230 (2010).

\bibitem{Pittman1995}
T. B. Pittman, Y. H. Shih, D. V. Strekalov, and A. V. Sergienko, 
{\em Optical imaging by means of two-photon quantum entanglement,} 
   Phys. Rev. A {\bf 52}, R3429--R3432 (1995).


























\bibitem{Florin1997}
E.-L. Florin, A. Pralle,  J. K. H. H{\"o}rber, and E. H. K. Stelzer, 
{\em Photonic force microscope based on optical tweezers and two-photon excitation for biological applications,} 
  J. Struct. Biol. {\bf 119}, 202--211 (1997).

\bibitem{Ghislain1993}
L. P. Ghislain and W. W. Webb, 
{\em Scanning-force microscope based on an optical trap,} 
  Opt. Lett. {\bf 18}, 1678--1680 (1993).


\bibitem{Friese1999}
M. E. J. Friese, A. G. Truscott, H. Rubinsztein-Dunlop, and N. R. Heckenberg,	
 {\em Three-dimensional imaging with optical tweezers,} 
  Appl. Opt. {\bf 38}, 6597--6603 (1999).


\bibitem{Tischer2001}
C. Tischer, S. Altmann, S. Fi{\u s}inger, J. K. H.  H{\"o}rber, E. H. K. Stelzer, and  E.-L. Florin, 
{\em Three-dimensional thermal noise imaging,} 
  Appl. Phys. Lett. {\bf 79}, 3878--3880 (2001).


\bibitem{Pralle1998}
A. Pralle, E.-L. Florin, E. H. K. Stelzer, and  J. K. H. H{\"o}rber,
{\em Local viscosity probed by photonic force microscopy,} 
  Appl. Phys. A Mater. Sci. Process. {\bf 66}, S71--S73 (1998).


\bibitem{Scholz2005}
T. Scholz, S. M. Altmann, M. Antognozzi, C. Tischer, J. K. H. H{\" o}rber, and B. Brenner,  
 {\em Mechanical properties of single myosin molecules probed with the photonic force microscope,}   
  Biophys. J. {\bf 88}, 360--371 (2005).
 
 
\bibitem{Rohrbach2004}
A. Rohrbach, C. Tischer, D. Neumayer, E. L. Florin, and  E. H.  Stelzer, 
{\em Trapping and tracking a local probe with a photonic force microscope,} 
  Rev. Sci. Instrum. {\bf 75}, 2197--2210 (2004).


 
 
  \bibitem{Bertseva2009}
E.  Bertseva {\it et al.}, 
{\em Intracellular nanomanipulation by a photonic-force microscope with real-time acquisition of a 3D stiffness matrix,}  
 Nanotechnology {\bf 20}, 285709 (2009). 











\bibitem{Peterman2003}
E. J. G. Peterman, F. Gittes, and  C.~F. Schmidt, 
{\em Laser-induced heating in optical traps,} 
  Biophys. J.{ \bf 84}, 1308--1316 (2003).

\bibitem{Neuman1999}
K.~C. Neuman, E.~H. Chadd, G.~F. Liou, K. Bergman, and S.~M. Block, 
{\em Characterization of photodamage to Escherichia coli in optical traps,} 
  Biophys. J. { \bf 77}, 2856–-2863 (1999).


\bibitem{Lubart2006}
R. Lubart, R. Lavi, H. Friedmann, and S. Rochkind, 
{\em Photochemistry and photobiology of light absorption by living cells,} 
  Photomed. Laser Surg.  {\bf 24}, 179--185 (2006). 





\bibitem{Fei1997}
 H.-B. Fei,  B. M. Jost, S. Popescu, B. E. A. Saleh, and  M. C. Teich, 
 {\em Entanglement-induced two-photon transparency,} 
  Phys. Rev. Lett. {\bf 78}, 1679 (1997).










\bibitem{Lopaeva2013}
E. D. Lopaeva, I. R. Berchera, I. P. Degiovanni, S. Olivares, G. Brida, and M. Genovese,  
{\em Experimental realization of quantum illumination,} 
    Phys. Rev. Lett. {\bf 110}, 153603 (2013). 




\bibitem{Boyer2008}
 V. Boyer, A. M. Marino, R. C. Pooser, and  P. D. Lett, 
{\em  Entangled images from four-wave mixing,} 
 Science {\bf 321}, 544--547 (2008). 




\bibitem{Mosset2005}
 A. Mosset, F. Devaux, and  E. Lantz, 
{\em  Spatially noiseless optical amplification of images,} 
  Phys. Rev. Lett. {\bf 94}, 223603 (2005).



\bibitem{Lopez2008}
L. Lopez,  N. Treps, B. Chalopin, C. Fabre, and  A. Ma\^{i}tre, 
{\em Quantum processing of images by continuous wave optical parametric amplification,} 
  Phys. Rev. Lett. {\bf 100}, 013604 (2008). 






\bibitem{Lassen2007}
M. Lassen, V. Delaubert, J. Janousek, K. Wagner, H.-A. Bachor, P. K. Lam, N. Treps, P. Buchhave, C. Fabre, and C. C. Harb,
{\em  Tools for multimode quantum information: Modulation, detection, and spatial quantum correlations,} 
  Phys. Rev. Lett. {\bf 98}, 083602 (2007).





\bibitem{Armstrong2012}
S. Armstrong, J.-F. Morizur, J. Janousek, B. Hage, N. Treps, P. K. Lam, and H.-A. Bachor, 
{\em Programmable multimode quantum networks,} 
  Nat. Commun. {\bf 3}, 1026 (2012).







\bibitem{Treps2003}
 N. Treps, N. Grosse, W. P. Bowen, C. Fabre, H.-A. Bachor, P. K. Lam, 
{\em A quantum laser pointer,}  
   Science {\bf 301}, 940--943 (2003).




\bibitem{Janousek2009}
J. Janousek, K. Wagner, J. F. Morizur, N. Treps, P. K. Lam, C. C. Harb, and H.-A. Bachor, 
{\em Optical entanglement of co-propagating modes,} 
  Nat. Photon. {\bf 3}, 399--402 (2009).













\bibitem{Selhuber2009}
C. Selhuber-Unkel, P. Yde, K.  Berg-S{\o}rensen, and  L.~B.~K.  Oddershede, 
{\em Variety in intracellular diffusion during the cell cycle,} 
 Phys. Biol. {\bf 6}, 025015 (2009).


\bibitem{Taylor2013_sqz}
M.~A. Taylor, J.~Janousek, V.~Daria, J.~Knittel, B.~Hage, H.-A.~Bachor, and W.~P.~Bowen,
{\em Biological measurement beyond the quantum limit,} 
  Nat. Photon. {\bf 7}, 229–-233 (2013).






\bibitem{McKenzie2004}
K. McKenzie, N. Grosse, W. P. Bowen, S. E. Whitcomb, M. B. Gray, D. E. McClelland, and P. K. Lam, 
 {\em Squeezing in the audio gravitational-wave detection band,} 
  Phys. Rev. Lett. {\bf 93}, 161105  (2004).


\bibitem{Taylor2013_lockin}
M. A. Taylor, J. Knittel, and W. P.  Bowen, 
{\em Optical lock-in particle tracking in optical tweezers,}
 Opt. Express {\bf 21}, 8018-–8024 (2013).





\bibitem{Kudo1990}
S. Kudo, Y. Magariyama, and  S.~I. Aizawa,  
{\em Abrupt changes in flagellar rotation observed by laser dark-field microscopy,} 
 Nature {\bf 346} 677--680 (1990).




\bibitem{Taylor2013darkfield}
M. A. Taylor and W. P.  Bowen, 
{\em Enhanced sensitivity in dark-field microscopy by optimizing the illumination angle,} 
 Appl. Opt. {\bf 52}, 5718-–5723 (2013).


\bibitem{Norrelykke2004}
I. M. Toli\'{c}-N{\o}rrelykke, E.-L. Munteanu, G. Thon, L. Oddershede, and K. Berg-S{\o}rensen, 
{\em Anomalous diffusion in living yeast cells.} 
 Phys. Rev. Lett. { \bf 93}, 078102  (2004).



\bibitem{Gittes1997}
F. Gittes, B. Schnurr, P.~D. Olmsted, F. C. MacKintosh, and C. F. Schmidt,  
{\em  Microscopic viscoelasticity: shear moduli of soft materials determined from thermal fluctuations,} 
 Phys. Rev. Lett. {\bf 79}, 3286--3289 (1997). 

\bibitem{Mason1997}
T. G. Mason, K. Ganesan, J. H. Van Zanten, D. Wirtz, and S. C. Kuo,  
 {\em Particle tracking microrheology of complex fluids,} 
  Phys. Rev. Lett. {\bf 79}, 3282--3285 (1997).






\bibitem{Weiss2004crowding}
M. Weiss, M. Elsner, F. Kartberg, and  T. Nilsson,  
 {\em Anomalous subdiffusion is a measure for cytoplasmic crowding in living cells,} 
 Biophys J. {\bf 87}, 3518--3524 (2004).













\bibitem{Stefszky2012}
M. S. Stefszky, C. M. Mow-Lowry, S. S. Y. Chua, D. A. Shaddock, B. C. Buchler, H. Vahlbruch, A. Khalaidovski, R. Schnabel, P. K. Lam, and D. E. McClelland, 
{\em Balanced homodyne detection of optical quantum states at audio-band frequencies and below,} 
 Class. Quantum Grav. {\bf 29}, 145015 (2012).

\bibitem{Mehmet2012}
M. Mehmet, H. Vahlbruch, N. Lastzka, K. Danzmann, and R. Schnabel, 
 Observation of squeezed states with strong photon-number oscillations, 
{\em Phys. Rev. A} {\bf 81}, 013814 (2010).


\bibitem{Taylor2013QNL}
M. A. Taylor, J. Knittel, and W. P.  Bowen, 
{\em Fundamental constraints on particle tracking with optical tweezers,}  
  New J. Phys. {\bf 15}, 023018 (2013).


\bibitem{Tsang2009}
M. Tsang, 
{\em Quantum imaging beyond the diffraction limit by optical centroid measurements,} 
   Phys. Rev. Lett. {\bf 102}, 253601 (2009).













\bibitem{Guigas2008}
G. Guigas and M. Weiss,  
{\em Sampling the cell with anomalous diffusion--the discovery of slowness,}  
 Biophys. J. {\bf 94}, 90--94 (2008).

\bibitem{Frauenfelder2006}
H. Frauenfelder, P. W. Fenimore, G. Chen, and  B. H. McMahon, 
{\em Protein folding is slaved to solvent motions,} 
 Proc. Natl. Acad. Sci. USA {\bf 103}, 15469--15472 (2006).



\bibitem{Tseng2004}
Y. Tseng, J. S. Lee, T. P. Kole, I. Jiang, and D. Wirtz,  
{\em Micro-organization and visco-elasticity of the interphase nucleus revealed by particle nanotracking,} 
 J. Cell Sci. {\bf 117}, 2159--2167 (2004).



\bibitem{Kucsko2013}
G. Kucsko, P. C. Maurer, N. Y. Yao, M. Kubo, H. J. Noh, P. K. Lo, H. Park, and M. D. Lukin,
{\em Nanometre-scale thermometry in a living cell,} 
 Nature {\bf 500}, 54-–59 (2013).


\bibitem{LeSage2013}
D. Le Sage, K. Arai, D. R. Glenn, S. J. DeVience, L. M. Pham, L. Rahn-Lee, M. D. Lukin, A. Yacoby, A. Komeili, and R. L. Walsworth, 
{\em Optical magnetic imaging of living cells,} 
 Nature {\bf 496}, 486--489 (2013).


\bibitem{Steinert2013}
S. Steinert, F. Ziem, L. T. Hall, A. Zappe, M. Schweikert, N. G{\" o}tz, A. Aird, G. Balasubramanian, L. Hollenberg, and J. Wrachtrup, 
{\em Magnetic spin imaging under ambient conditions with sub-cellular resolution,} 
 Nat. Commun. {\bf 4}, 1607 (2013).


\bibitem{Kaufmann2013}
S. Kaufmann {\it et al.}, 
 {\em Detection of atomic spin labels in a lipid bi-layer using a single-spin nanodiamond probe,}  
 Proc. Natl. Acad. Sci. USA {\bf 110}, 10894–-10898 (2013).





\end{thebibliography}
\end{document}